\documentclass[prc,aps,twocolumn,floats,showpacs,superscriptaddress,byrevtex]{revtex4}
\input epsf.tex
\newcommand{\etal}{\emph{et al.}}
\begin{document}
\title{Shape Coexistence in $^{186}$Pb:  
       Beyond-mean-field description\\ 
       by configuration mixing of symmetry restored wave functions}
\author{T. Duguet}
\altaffiliation[Current address: ]{Physics Division, 
                Argonne National Laboratory, Argonne, IL 60439}
\affiliation{Service de Physique Th\'eorique, 
             CEA Saclay, 91191 Gif sur Yvette Cedex, 
             France}
\author{M. Bender}
\affiliation{Service de Physique Nucl\'eaire Th\'eorique, 
             Universit\'e Libre de Bruxelles, C.P. 229, B-1050 Bruxelles, 
             Belgium}
\author{P. Bonche}
\affiliation{Service de Physique Th\'eorique, 
             CEA Saclay, 91191 Gif sur Yvette Cedex, 
             France}
\author{P.-H. Heenen}
\affiliation{Service de Physique Nucl\'eaire Th\'eorique, 
             Universit\'e Libre de Bruxelles, C.P. 229, B-1050 Bruxelles, 
             Belgium}
\date{February 28 2003}
\begin{abstract}
We study shape coexistence in $^{186}$Pb using configuration mixing of 
angular-momentum and particle-number projected self-consistent
mean-field states. The same Skyrme interaction SLy6 is used everywhere 
in connection with a density-dependent, zero-range pairing force.
The model predicts coexisting spherical, prolate and oblate $0^+$ 
states at low energy.
\end{abstract}
\maketitle
%
%
The behavior of shell effects away from the valley of stability is a topic 
of very active investigation, both theoretically and experimentally. 
For light nuclei, the spherical \mbox{$N=20$} and \mbox{$N=28$} shells 
disappear in neutron-rich isotopes, leading to strongly deformed ground 
states and large $B(E2)$ transition probabilities between the first 
$2^+$ state and the ground state \cite{Rei99}. 
In contrast, the magic proton number \mbox{$Z=82$} 
is particularly strong and its influence persists even in very 
neutron-deficient nuclei. The ground state of Pb isotopes is known to
be spherical down to $^{182}$Pb \cite{Wau94}. The weakening of the 
magicity of the \mbox{$Z=82$} shell manifests itself through the appearance 
of low-lying $0^+$ states \cite{Jul01}. At least one low-lying, excited 
$0^+$ level has been observed in all even-even Pb isotopes between 
\mbox{$A=182$} and 194 at excitation energies below 1 MeV,
the most extreme cases being $^{188}$Pb and $^{186}$Pb 
\cite{Hee93,And00} with two excited $0^+$ states below 700 keV. 

Two different kinds of models have been invoked to explain the
coexistence of several $0^+$ states at low energy \cite{Woo92}. 
In a shell model picture \cite{And00}, the first excited 
$0^+$ level observed from $^{202}$Pb down to $^{186}$Pb is interpreted 
as a two-quasiparticle proton configuration $(\pi h9/2)^2$, while 
the second one in $^{188}$Pb and $^{186}$Pb as well as the first $0^+$ 
state in $^{184}$Pb are understood as a four-quasiparticle configuration 
$(\pi h9/2)^4$. In this picture, neutrons and protons outside the 
inert core interact through pairing and quadrupole interactions 
to generate deformed structures. Such a model requires a 
drastic truncation of the configuration space. Up to now, it has 
only been applied in rather schematic and qualitative ways.

In mean-field models, the $0^+$ states observed at low energies
are associated with coexisting energy minima which
appear for different values of the axial quadrupole moment \cite{naz93}. 
The ground state corresponds to the spherical minimum and the excited 
$0^+$ level to a deformed state with an oblate (in the heaviest Pb 
isotopes) or a prolate (in $^{184}$Pb up to $^{188}$Pb) shape.

However, shape coexistence in the neutron-deficient Pb region 
cannot be described on the level of mean-field models in a fully 
satisfactory way. The minima obtained as a function of the 
quadrupole moment are rather shallow and dynamical effects such 
as quadrupole vibrations may affect the very existence of 
these minima. Tajima \etal\ \cite{Taj93} and, more recently, Chasman 
\etal\ \cite{cha01} have studied the quadrupole dynamics of Pb isotopes 
by performing a configuration mixing of mean-field states with
different axial quadrupole moments. Their results support the 
interpretation of the excited $0^+$ states as deformed minima. 
The lowest excited levels obtained in the configuration mixing 
calculation have average deformations close to that of 
the mean-field minima. However, the calculated excitation energies 
overestimate the experimental values. Diabatic effects have been 
studied by Tajima \etal\ who have included, for each axial 
quadrupole moment, the lowest Hartree-Fock+BCS (HFBCS) configuration 
and the two-quasiparticle deformed proton configurations 
$(\pi h9/2)^2$. Tajima \etal\ have shown that these configurations 
do not influence the configuration-mixing results significantly,
and that they can be neglected. 

The experimental data on neutron deficient Pb isotopes are 
not limited to a few $0^+$ states. Rotational bands have also been 
observed whose properties have served to interpret the excited $0^+$ 
state as associated with oblate and prolate deformations. Transition 
probabilities between the levels are also known in some cases. 
It seems, therefore, highly desirable to apply the configuration-mixing 
method that we have recently developed \cite{Val01} to Pb isotopes. 
This method treats simultaneously the most important symmetry 
restorations and the mixing with respect 
to a collective variable. Here, we present an application to 
$^{186}$Pb. This isotope has the unique property of having 
$0^+$ levels as its lowest three states, with the excitation 
energy of the second and third $0^+$ also being the lowest among the 
known Pb isotopes \cite{And00}. While the ground state is assumed 
to be spherical, the $0^+$ states observed at 532 keV and 650 keV 
are interpreted as corresponding to oblate and prolate configurations.

The ``projected'' configuration mixing of mean-field wave functions 
performed here has several goals. The particle-number projection removes 
unwanted contributions coming from states with different particle 
numbers, which are an artifact of the BCS approach. The angular
momentum projection separates the contribution from different
angular momenta to the mean-field states and generates wave functions 
in the laboratory frame with good angular momentum. Finally, 
the variational configuration mixing with respect to a collective
coordinate, the axial quadrupole moment in this work, removes the
contributions to the ground state coming from collective vibrations,
and simultaneously provides the excitation spectrum corresponding 
to this mode.

The starting point of our method is a set of independent HFBCS 
wave functions $| q \rangle$ generated by mean-field calculations 
with a constraint on a collective coordinate $q$. Such mean-field 
states break several symmetries of the exact many-body states.
Wave functions with good angular momentum and particle numbers 
are obtained by the restoration of rotational and particle-number 
symmetry on $ | q \rangle$:
\begin{equation}
\label{eq:proj}
|J M q \rangle 
= \frac{1}{{\mathcal N}}
  \sum_{K} g^{J}_{K}
  \hat{P}^J_{MK} \hat{P}_Z \hat{P}_N | q \rangle
,
\end{equation}
where ${\mathcal N}$ is a normalization factor. 
$\hat{P}^{J}_{MK}$, $\hat{P}_N$, $\hat{P}_Z$ are projectors 
onto the angular momentum $J$ with projection $M$ along the laboratory 
z-axis, neutron number $N$ and proton number $Z$, respectively. 
We impose axial symmetry and time reversal invariance and, therefore,
$K$ can only be 0 and we shall omit the coefficient 
\mbox{$g^{J}_{K} = \delta_{K0}$}. This prescription excludes the 
description of $\gamma$ bands where \mbox{$K=2$}.

A variational configuration mixing on the collective variable 
$q$ is then performed for each angular momentum
\begin{equation}
\label{eq:discsum}
| J M k \rangle 
= \sum_{q} f_{k}^{JM} (q) | JM q \rangle 
.
\end{equation}
The weight functions $f_k^{JM}(q)$ are determined by requiring
that the expectation value of the energy
\begin{equation}
\label{eq:egcm}
E^{JM}_k
= \frac{\langle JM k | \hat H | JM k \rangle}
       {\langle JM k | JM k\rangle}
\end{equation}
is stationary with respect to an arbitrary variation $\delta f_k^{JM}(q)$.
This prescription leads to the discretized Hill-Wheeler equation \cite{HW53}. 
Such a secular problem amounts to a restricted variation after projection 
in the set of states obtained for different values of the collective 
variable $q$. Collective wave functions in the basis of 
the intrinsic states are then obtained from the set of weight 
functions $f_k^{JM}(q)$ by a basis transformation \cite{Taj93}.
In $| J M k \rangle$, the weight of each mean-field state $| q \rangle$
is given by~:
\begin{equation}
\label{eq:weight}
g^{JM}_k(q)
= {\langle JM k | q \rangle}
.
\end{equation}
Since the collective states $| JM k \rangle$ have good angular 
momentum, quadrupole moments and transition probabilities can 
be determined directly in the laboratory frame of reference without
further approximations.

The same effective interaction is used to generate the mean-field wave 
functions and to perform the configuration mixing calculation. We have chosen 
the Skyrme interaction SLy6 in the mean-field channel \cite{Cha98} 
and a density-dependent, zero-range force as defined in \cite{Rig99},
in the pairing channel. The pairing equations are solved using 
the Lipkin-Nogami prescription, as done in \cite{Val01}.
The two-body center-of-mass correction is self-consistently included 
in the interaction SLy6 \cite{Cha98}. However, in the present calculations, 
it is included \emph{a posteriori} at the mean-field level as well as in 
the projection and configuration-mixing calculations.
%
%

%
%
\begin{figure}[t!]
\epsfbox{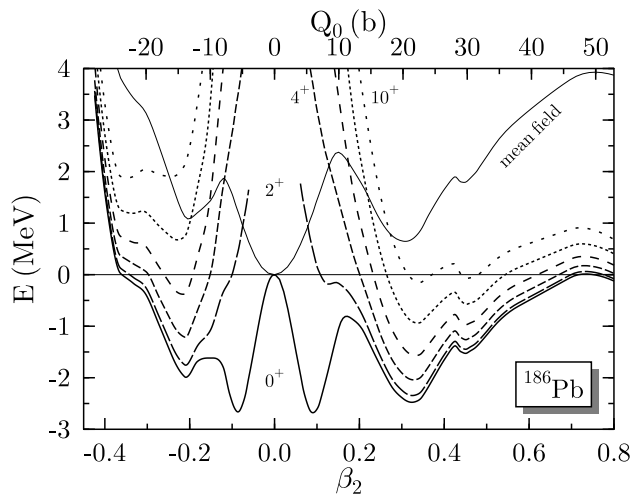}
\caption{\label{fig:eproj}
Particle-number projected (``mean field'') and particle-number
and angular-momentum projected potential energy curves up to 
\mbox{$J=10$} for $^{186}$Pb as a function of the mass 
quadrupole moment in barn (upper axis) or, equivalently, in terms 
of $\beta_2$ (lower axis). The energy reference is that of the 
projected spherical mean-field state.
}
\end{figure}
%
%

In figure \ref{fig:eproj} the deformation energy of $^{186}$Pb 
is plotted before and after projection on angular momentum.
All curves are drawn versus the intrinsic axial quadrupole moment 
of the unprojected mean-field states. As projected \mbox{$J=0$} 
states are spherical, this ``quadrupole moment'' is only a 
convenient way to label the projected states. The curve labeled 
``mean-field'' plots the deformation energy after particle-number 
projection only. It exhibits a spherical global minimum as well 
as local minima at prolate and oblate deformations.
While the deformation energy of the prolate minimum fortuitously
reproduces the experimental value of 0.650 MeV for the prolate 
$0^+$ state, the 1.1 MeV deformation energy of the oblate minimum 
overestimates the experimental value of 0.532 MeV for 
the oblate $0^+$ level. A fourth, very shallow, minimum can 
be seen at a deformation \mbox{$\beta_2 \approx 0.5$}; it is too 
shallow to be safely associated with a physical state. 

The energy curves obtained after angular momentum projection 
are also shown in figure \ref{fig:eproj}. At moderate deformations, 
around the prolate and oblate minima, the mean-field states are 
dominated by angular momentum components with \mbox{$J \geq 8$}.
This is reflected in the fact that all projected energy curves 
are far below the mean-field one. The spherical mean-field state
is rotationally invariant and, therefore, contributes to \mbox{$J=0$} 
only. Two minima appear at small deformations, around 
\mbox{$\beta_2 = \pm 0.1$}. They do not correspond to two different 
states, but to the correlated spherical state (see below). 
For larger prolate and oblate deformations, the energy difference 
between the mean-field and \mbox{$J=0$} curves stays nearly constant.
The prolate and oblate mean-field minima are present in all 
the projected energy curves. Angular momentum projection reduces
the energy difference between the spherical ($|\beta_2| \approx 0.1$) 
and deformed minima to 0.2 MeV for the prolate and 0.68 MeV for 
the oblate well. While the prolate potential well is pronounced 
for all angular momenta, the oblate one now becomes very shallow 
for \mbox{$J=0$}. 

%
%
\begin{figure}[t!]
\epsfbox{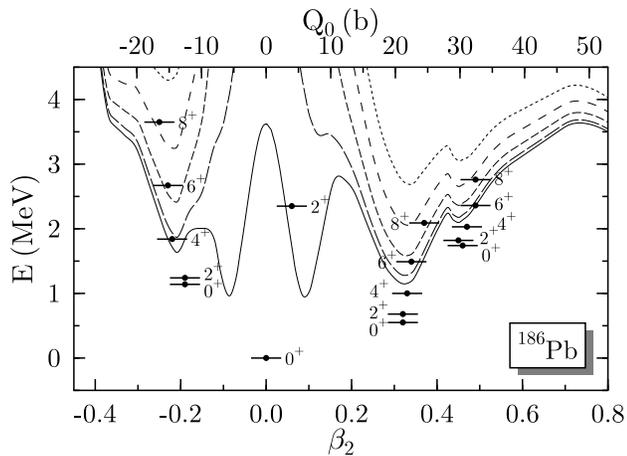}
\caption{\label{fig:ex}
Spectrum of the lowest positive parity bands with even angular momentum 
and \mbox{$K=0$}, as a function of the deformation (see text).
The angular momentum projected energy curves are shown for comparison. 
The energy reference is that of the calculated $0^+_1$ ground state. 
}
\end{figure}
%
%
The excitation energies $E^{JM}_k$ of the collective states 
$| JM k \rangle$ obtained from the configuration mixing calculation
are shown in figure \ref{fig:ex}. Each of these states is 
represented by a horizontal bar drawn at the average intrinsic 
deformation $\sum_q \beta_2(q) \, |g_k^{JM}(q)|^2$, where 
$\beta_2(q)$ is the deformation of the mean-field state. 
The excitation spectrum is divided into bands associated with 
different deformations. Configuration mixing lowers the energy of 
the lowest collective states with respect to the projected energy 
curves. The energy gain is the largest for the ground state, hence
increasing the excitation energies of the prolate and oblate 
$0^+$ levels.
%
%
\begin{figure}[b!]
\epsfbox{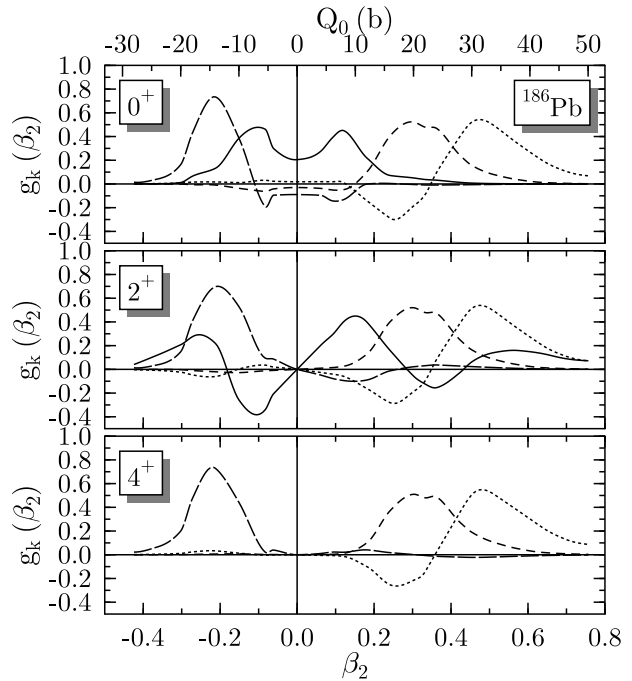}
\caption{\label{fig:wf}
GCM wave functions of the lowest $| J k \rangle$ states. Solid 
lines denote spherical, long-dashed lines oblate, dashed lines prolate, 
and dotted lines the $\beta$ band in the prolate well.
}
\end{figure}
%
%

The corresponding collective wave functions $g^{JM}_k(q)$ are 
presented in figure \ref{fig:wf}. Their square gives the weight 
of each mean-field state $| q \rangle$ in the collective state 
$| JM k \rangle$. 

The ground state wave function is spread in a similar way on both 
oblate and prolate sides with a zero value for the average $\beta_2$ 
deformation. The wave functions of the first two excited $0^+$ states are 
strongly peaked at either prolate ($0^+_2$) or oblate deformation 
($0^+_3$), with their tails extending into the spherical well. 
For higher $J$ values, the shape of the wave functions confirms 
their assignment to oblate and prolate bands, as already hinted 
so in figure \ref{fig:ex}. As there is no spherical well for 
\mbox{$J>0$} states, their wave functions mix only prolate and 
oblate configurations. Starting with \mbox{$J=4$}, all levels are 
strongly localized and are predominantly either prolate or oblate.
The shapes of the $0^+_4$, $2^+_3$, and $4^+_3$ wave functions 
suggest their interpretation as a rotational band built on a $\beta$ 
vibration within the prolate well, while the wave function of the $2^+_4$ 
state indicates that it corresponds to a vibrational state, which is spread 
over the entire potential well.

\begin{figure}[t!]
\epsfbox{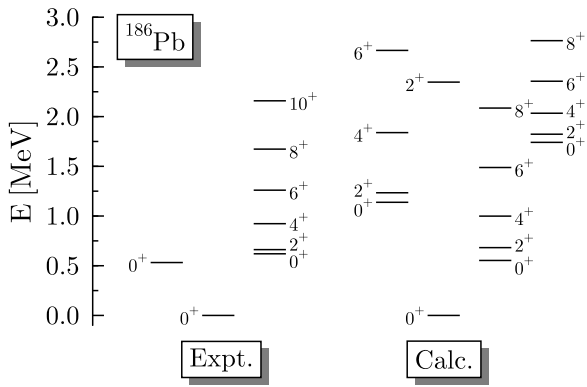}
\caption{\label{fig:spect}
Comparison between the calculated excitation energies and the available
experimental data for low-lying states in $^{186}$Pb. From the left to
the right the spectra show oblate, spherical and prolate bands.
}
\end{figure}

The calculated spectrum is compared with the experimental data in figure 
\ref{fig:spect}. The excitation energy of the prolate $0^+$ state at 0.55 MeV 
is very close to the experimental value. In contrast, the excitation 
energy of the oblate $0^+$ level is largely overestimated. The 
experimental data even suggest that the oblate state is slightly below 
the prolate one. A nice result from the calculations is that the structure 
of the first three $0^+$ levels is dominated by spherical, prolate and 
oblate configurations, respectively, and this supports the interpretation 
of the experimental data in terms of shape coexistence. This feature could 
not have been guessed from the deformation energy curves (see figure 
\ref{fig:eproj}), where the oblate well has a depth of only 500 keV in 
contrast with the prolate one of 1 MeV. The excitation energies of the first 
two excited $0^+$ states are even quite close to the energy differences 
between the deformed minima and the spherical minimum of the mean-field 
deformation energy curve.
%
%
%
%

Both experimentally and theoretically, all bands exhibit a rotational 
behavior, with the exception of  the $E_{0^+}-E_{2^+}$ energy difference
which is too small. This can be understood from the stronger state 
mixing for the \mbox{$J=0$} than for higher $J$ values which is 
observed in the calculations.
For the prolate band, however, the displacement from a rotational behavior 
remains too small. This is probably a consequence of the overestimated energy 
of the oblate band head which reduces the mixing between the deformed 
configurations.

Calculated transition probabilities for \mbox{$J>2$} states confirm 
the separation of the excited states into rotational bands with very small 
$B(E2)$ transitions between them. While the transition quadrupole moments, 
$Q_0$, of the oblate ($Q_0\approx -600$ $e$ fm$^2$ or $\beta_2\approx -0.2$)
and prolate ($Q_0\approx 1000$ $e$ fm$^2$ or $\beta_2\approx 0.34$)
bands slowly grow with angular momentum, the deformation of the 
third rotational band stays nearly constant at about 
$Q_0\approx 1400$ $e$ fm$^2$ ($\beta_2\approx 0.49$), in
agreement with the systematics of the minima in the projected energy 
curves of figure \ref{fig:eproj}. The $B(E2)$ values for the in 
and out of band $2^+ \to 0^+$ transitions confirm that the 
low-lying $0^+$ states are indeed mixed. 

Our results strongly support the interpretation of the Pb isotopes 
spectra as evidence for shape coexistence. There remains, however,
a significant over-estimation of one of the band's excitation energy.
This could be due to several ingredients of the model:

$\bullet$ the effective mean-field interaction; small differences 
between interactions (surface tension, spin-orbit strength \dots)
shift the relative energies of the various coexisting minima
at the mean-field level~\cite{Ben02,Nik02}. 
In a calculation with the Skyrme SLy4 interaction,
the prolate and oblate $0^+$ states are pushed up to 1.05 MeV and 
1.39 MeV, respectively, as can be expected from the overall stiffer 
energy surface of this interaction \cite{Ben00}.

$\bullet$ the strength and the form factor of the pairing interaction; 
a test with a reduced pairing strength ($-1100$ MeV fm$^3$) shows that 
the energies of the prolate and oblate minima of the deformation energy 
curves are reduced to 0.2 MeV and 0.65 MeV, respectively.

$\bullet$ the configuration space used in the configuration mixing; 
to test this possible source of error, we have enlarged the space
by including the oblate $(\pi h9/2)^2$ two-quasiparticle proton 
configurations, as was done by Tajima~\etal\ \cite{Taj93}. 
As in this work, the results are changed by at most 100 keV.

$\bullet$ the inclusion of triaxial quadrupole configurations;
projection on $J$ becomes much heavier numerically, and this is 
still beyond present numerical possibilities. 

$\bullet$ generalized interaction for calculations beyond mean field;
most mean-field interactions depend on the one-body density.
It is known since the 70s that this density dependence can have 
two different origins: either a three-body force or a resummation 
of (short-range) correlations. To generalize 
this dependence for the non-diagonal matrix elements appearing 
beyond the mean-field approximation, we have chosen the 
generalisation stemming from a three-body interaction. Resummation of 
correlations beyond mean-field gives rise to another generalisation 
of the Skyrme force \cite{Dug02}. The study of shape coexistence 
in nuclei like the Pb isotopes could be a good place to
determine the merits of both generalisations of the Skyrme force. 
%
%
\begin{acknowledgments}
This research was supported in part by the PAI-P5-07 of the Belgian
Office for Scientific Policy. We thank M.~Huyse, R.~Janssens 
and P.\ Van Duppen for fruitful and inspiring discussions.
M.~B.\ acknowledges support through a European Community Marie Curie
Fellowship. 
\end{acknowledgments}
%
%

\end{document}